\documentclass[aps,prd,10pt,twocolumn,nofootinbib]{revtex4}

\usepackage{epsfig}
\usepackage{bm}
\usepackage{latexsym}
\usepackage{natbib}
\usepackage{url}
\usepackage{dcolumn}
\usepackage{color}
\usepackage{amsfonts,amssymb,amsmath}
\usepackage{graphicx,epsfig}
\usepackage{psfrag}
\usepackage{subfigure}
\usepackage{hyperref}
\hypersetup{colorlinks=true}
\usepackage{mathtools}
\usepackage{enumitem}
\usepackage{float}
\usepackage[dvipsnames]{xcolor}

\usepackage{mathrsfs}

\newcommand{\scrip}{\mathscr{I}^{+}}
\newcommand{\scrim}{\mathscr{I}^{-}}

\newcommand{\zbr}{\bar{z}}

\def\IZ{{\mathbb Z}}

\def\be{\begin{equation}}
\def\ee{\end{equation}}
\def\beq{\begin{equation}}
\def\eeq{\end{equation}}
\def\bea{\begin{eqnarray}}
\def\eea{\end{eqnarray}}

\begin{document}

\title{A New Gauge for Asymptotically Flat Spacetime}
\author{
Chethan Krishnan$^{a}$,  Jude Pereira$^{b}$ \\
\vspace{0.7cm}
}


\affiliation{$^a$ Centre for High Energy Physics, Indian Institute of Science, Bangalore 560012, India\\
{\rm Email:} \textmd{chethan.krishnan@gmail.com} 
\vspace{0.2cm}\\
$^b$ Department of Physics, Arizona State University, Tempe, Arizona  85287-1504, USA \\
{\rm Email:} \textmd{jude.pereira@asu.edu}
}

\begin{abstract}

We present a new gauge for asymptotically flat spacetime that can treat future and past null infinities ($\mathscr{I}^{+}$ or $\mathscr{I}^{-}$) democratically. Our gauge is complementary to Bondi and Ashtekar-Hansen gauges, and is adapted to the $S$-matrix being the natural observable. One new feature is that the holographic directions are null. We present a set of consistent fall-offs in terms of null coordinates at $\mathscr{I}^{+}$ and $\mathscr{I}^{-}$, with finite BMS$^{\pm}$ charges. The diagonal BMS$^0$ symmetry of the gravitational $S$-matrix emerges upon demanding {\em asymptotic} CPT invariance. Trivial diffeomorphisms, (absence of) log fall-offs, possible enhancements of BMS algebra, and the possibility of holographic renormalization of data at $\mathscr{I}^{+}_-$ and $\mathscr{I}^{-}_+$, play interesting roles. Gory details of the various new technical features that emerge, are elaborated in a companion paper to this letter.


\end{abstract}

\maketitle

\section{Introduction}




Attempts at flat space holography \cite{Strominger} are usually phrased in the Bondi gauge \cite{BBM, Sachs}. Bondi gauge comes in two varieties, adapted either to $\scrip$ or $\scrim$, but not both \cite{AshtHans}. In this letter, we will present a new gauge for asymptotically flat spacetime that can treat $\scrip$ or $\scrim$ on an equal footing. 

We feel that this is of utility for a few different reasons. The natural observable of flat space holography is the S-matrix, which by definition is tied to {\em both} the past and the future \cite{S}. Bondi and collaborators on the other hand were motivated by the desire to characterize gravitational waves {\em from} isolated gravitating systems. This has a retarded correlator feel to it, and Bondi gauge reflects that. A gauge that is adapted to the S-matrix should have a more Feynman-like feel, and should speak to both the future and the past. It is also worth noting here that Strominger's identification  \cite{Strominger1} of a diagonal BMS symmetry for gravitational scattering in flat space, crucially relies on an antipodal matching between the future and the past.

In flat space, the codimension-1 conformal boundary (which is presumably a natural habitat of the hologram \cite{cod1}) has pieces {\em both} in the future and past. A Poincare transformation in the bulk affects both null boundaries simultaneously. It has also been noted in recent work \cite{ACD1, ACD2} that Asymptotic Causal Diamonds (ACD) which are causal diamonds whose vertices are anchored to the future and past null boundaries, are a useful ingredient in understanding the entanglement structure of flat space holography \cite{cod2}. Yet another motivation that will play a role in our work is that the asymptotic (ie., holographic) coordinate in Bondi gauge is the radial direction, even though we are trying to get to $\scrip$ (or $\scrim$). It is more natural in our view, to think of these boundaries as being approached along a {\em null} direction. 

Motivations are one thing, but we will find that this approach also leads to some new and surprising results. These will be summarized in the concluding section. The goal of the present letter is to outline the main results, the technical details can be found in the more elaborate companion paper \cite{Big}. 



\section{Special Double Null Gauge}

In this letter we will present a gauge that can treat either null boundary of flat space on an equal footing \cite{future-past}. 
The considerations in the Introduction hint that we should work with a double-null or light-front coordinate system. Double null coordinates are well-explored, see eg. \cite{Israel}. It is easy enough to see \cite{Big} that it amounts to setting 
\bea
g^{uu}=0, \  \ g^{vv}=0 \label{dn}
\eea
where $u$ and $v$ are to be understood as null coordinates. But
these are only two conditions. To get to a genuine gauge choice in four dimensions, we need two more conditions. A simple choice that respects the future-past ``symmetry" that we wish to retain is the Special Double Null (SDN) gauge,
\bea
g^{uA}=g^{vA} \label{sdn}
\eea
where the $A$ index is any of the angle directions on the sphere, and therefore generalizes to arbitrary dimensions. A convenient way to write the metric in this gauge is
\begin{widetext}
\bea
\label{doublenull} ds^2=-e^{\lambda}du\ dv +\Big(\frac{v-u}{2}\Big)^2\Omega_{AB}(dx^A-\alpha^A du-\alpha^A dv)(dx^B-\alpha^B du-\alpha^B dv) 
\eea
\end{widetext}
where $\lambda, \Omega_{AB}, \alpha^A$ are the (necessary) six independent functions. 

Let us clarify a trivial, but potentially confusing,  point. The symmetry of the gauge conditions under the exchange of $u$ and $v$ legs of the metric tensor does {\em not} mean that we are allowing only solutions that allow some kind of discrete spacetime symmetry. The dependence of the metric functions on $(u,v)$ can of course still be arbitrary, and indeed can break any such symmetry. At a later stage however, we will demand an {\em asymptotic} CPT invariance on the solution space.

\section{Fall-offs: Riemann and Einstein}

So far what we have done is very general, and can be accomplished in any metric. To specify that we are working with an asymptotically flat spacetime, we need to specify fall-off conditions on this metric at infinity around Minkowski space. 
Minkowski space in SDN gauge can be written as
\bea
ds^2=-du\ dv + 2 \Big(\frac{v-u}{2}\Big)^2 \gamma_{z\zbr}(z,\zbr)\ dz d\zbr \label{Mink}
\eea
where the 2-sphere metric is
\beq \gamma_{zz}(z,\zbr)=\gamma_{\zbr\zbr}(z,\zbr)=0, \quad \gamma_{z\zbr}(z,\zbr)=\frac{2}{(1+z\zbr)^2} \label{complexsphere}\eeq 
We will consider asymptotic fall-offs around Minkowski of the form
\begin{subequations}\label{polyfalloffs}
    \begin{align}
        \label{polylambda}
        \lambda(u,v,z,\zbr) &= \frac{\lambda_{1}(u,z,\zbr)}{v}+\frac{\lambda_{2}(u,z,\zbr)}{v^2}+O\big(v^{-3}\big) \\
        \label{polyOmegazz}
        \Omega_{zz}(u,v,z,\zbr) &= \frac{\mathcal{C}_{zz}(u,z,\zbr)}{v}+\frac{\mathcal{D}_{zz}(u,z,\zbr)}{v^2}+O\big(v^{-3}\big)\\
        \label{polyOmegazw}
        \Omega_{z\zbr}(u,v,z,\zbr) &= \gamma_{z\zbr}(z,\zbr)+\frac{\mathcal{D}_{z\zbr}(u,z,\zbr)}{v^2}+O\big(v^{-3}\big)\\
        \label{polyOmegaww}
        \Omega_{\zbr\zbr}(u,v,z,\zbr) &=\frac{\mathcal{C}_{\zbr\zbr}(u,z,\zbr)}{v}+\frac{\mathcal{D}_{\zbr\zbr}(u,z,\zbr)}{v^2}+O\big(v^{-3}\big)\\
        \label{polyalphaz}
        \alpha^z(u,v,z,\zbr) &= \frac{\alpha^{z}_{\ 3}(u,z,\zbr)}{v^3}+\frac{\alpha^{z}_{\ 4}(u,z,\zbr)}{v^4}+O\big(v^{-5}\big)\\
        \label{polyalphaw}
        \alpha^{\zbr}(u,v,z,\zbr) &= \frac{\alpha^{\zbr}_{\ 3}(u,z,\zbr)}{v^3}+\frac{\alpha^{\zbr}_{\ 4}(u,z,\zbr)}{v^4}+O\big(v^{-5}\big)
    \end{align}
\end{subequations}
We could try slightly more general fall-offs if we wish, by allowing non-trivial $C_{z\zbr}, \alpha^z_2, \alpha^{\zbr}_2$, and also allowing a non-trivial conformal factor on the sphere as permitted by Penrose \cite{Penrose}. Some of these features will be discussed in \cite{Big} where we will also consider log terms in great detail. Here our goal is simply to work with a minimal fall-off choice that connects with some of the familiar features of asymptotically flat gravity (like BMS symmetry \cite{BBM, Sachs, Flanagan}). 
  
Note that as promised, the holographic direction is null. We have presented future fall-offs, a similar expansion in $1/u$ adapted to the past ($u=-\infty$) can and will be used later. Let us emphasize that unlike in Bondi, where there are in fact two different looking gauges in the future and past, here two copies of the same gauge are getting expanded at the two different boundary charts.

Demanding asymptotic flatness amounts to setting Riemann $=0$ asymptotically \cite{Weyl}. When we do this, naively we may expect to be left with \eqref{Mink} order by order (up to supertranslations and perhaps super-rotations). Instead we find a new feature unseen in the previous gauges -- Riemann flatness allows towers of undetermined integration ``constants" in the fall-off coefficients that are purely functions of $z$ and $\zbr$. They arise in the coefficients of $\lambda$ and $\alpha^{A}$ \cite{lamD}. We will see that most of them correspond to the freedom for doing {\em trivial} asymptotic diffeomorphisms \cite{Subtlety}. 

The integration ``constants" persist when we demand Einstein equations asymptotically \cite{RiemFlat}. The technical reason is that unlike in other gauges, Einstein equations generically only determine the first $u$-derivatives of the higher order coefficients in $\lambda$ and $\alpha^{A}$ \cite{Exceptions}. A simple and paradigmatic example one can keep in mind is \bea \partial_{u}\alpha^{z}_{\ 3} &= -2\, \big(\gamma^{z\zbr}\big)^2 \, D_{z}\mathcal{C}_{\zbr\zbr} \eea which leads to an integration ``constant" in $\alpha^{z}_{\ 3}$. We put ``constant'' in quotations to emphasize that it has angle dependence.

Apart from these, there are also physical data that determine the solution asymptotically. These can be chosen to be in the fall-off coefficients of
\bea
\mathcal{C}_{zz}, \ \mathcal{C}_{\zbr\zbr}, \ \lambda_2, \ \alpha^z_4, \ \alpha^{\zbr}_4.
\eea 
These are analogous to the two {\em shear data}, the {\em mass aspect} and the two {\em angular momentum aspect} data. We also define \cite{newsdef} 
\bea
\mathcal{N}_{AB}=\partial_u  \mathcal{C}_{AB}. \eea
The shear is (essentially) free data as it is in Bondi \cite{shear}, but the mass and angular momentum aspects satisfy the constraints:
\bea\partial_u^2\lambda_2=&\hspace{-0.35in}-\frac{1}{2}(\gamma^{z\zbr})^2D_z^2\mathcal{N}_{\zbr\zbr}-\frac{1}{2}(\gamma^{z\zbr})^2D_{\zbr}^2\mathcal{N}_{zz}+ \nonumber\\ &-\frac{1}{8}(\gamma^{z\zbr})^2\mathcal{C}_{\zbr\zbr}\, \partial_u\mathcal{N}_{zz}-\frac{1}{8}(\gamma^{z\zbr})^2\mathcal{C}_{zz}\, \partial_u\mathcal{N}_{\zbr\zbr},\label{lambda2eq}\eea
\bea
    \partial^2_u\alpha^z_4 =& \hspace{-0.35in}-2\gamma^{z\zbr} D_{\zbr}(\partial_u\lambda_2)-\frac{1}{2}(\gamma^{z\zbr})^3\mathcal{N}_{zz}\, D_{\zbr}\mathcal{C}_{\zbr\zbr}+\nonumber\\&+\frac{7}{2}(\gamma^{z\zbr})^3\mathcal{N}_{\zbr\zbr}\, D_{\zbr}\mathcal{C}_{zz}+\nonumber\\ &+\frac{1}{2}(\gamma^{z\zbr})^3 \mathcal{C}_{\zbr\zbr}\, D_{\zbr}\mathcal{N}_{zz}+\frac{1}{2}(\gamma^{z\zbr})^3\mathcal{C}_{zz}\, D_{\zbr}\mathcal{N}_{\zbr\zbr}+\nonumber\\&-6u(\gamma^{z\zbr})^2 D_z\mathcal{N}_{\zbr\zbr}-4(\gamma^{z\zbr})^2D_z\mathcal{C}_{\zbr\zbr}+
\nonumber\\&+4(\gamma^{z\zbr})^3D_zD_{\zbr}D_z\mathcal{C}_{\zbr\zbr} +\nonumber\\&-2(\gamma^{z\zbr})^3 D_{\zbr}D^2_z\mathcal{C}_{\zbr\zbr}-2(\gamma^{z\zbr})^3D^3_{\zbr}\mathcal{C}_{zz},
\eea
together with a similar expression for $\alpha^{\zbr}_4$. The structures above have parallels to BMS; a key distinction to note is that unlike the analogous equations in Bondi, these are second derivative constraints. The {\em second} integration ``constants" here are trivial \cite{hardwire}, and the physical data is contained in $\partial_u \lambda_2$ and $\partial_u \alpha^{A}_{\ 4}$ -- these are the true analogues of the mass and angular momentum aspects. We have checked that Kerr is contained in this.
 

\section{Asymptotic Killing Vectors: Trivial Diffeomorphisms, BMS Algebra and Covariant Surface Charges}
With the malice of hindsight, we will look for asymptotic diffeomorphisms that preserve our gauge choice \eqref{doublenull} and the following fall-offs:
\begin{subequations}\label{boundarycond}
    \begin{align}
        g_{uu} &= g_{vv}= O\big(v^{-4}\big)\\
        g_{uv} &= -\frac{1}{2}+O\big(v^{-2}\big)\\
        g_{z\zbr} &= \frac{1}{4}\, \gamma_{z\zbr}\, v^2-\frac{1}{2}\, u\, \gamma_{z\zbr}\, v+O\big(v^{0}\big)\\
        g_{zz} &= O(v)\\
        g_{\zbr\zbr} &= O(v)\\
        g_{uA} &= g_{vA} = O\big(v^{-1}\big)
    \end{align}
\end{subequations}
Correspondingly, we demand the exact Killing equations for our gauge \bea \mathcal{L}_{\xi} g^{uu} = 0,\ \mathcal{L}_{\xi} g^{vv} = 0,\ \mathcal{L}_{\xi} g^{uA} = \mathcal{L}_{\xi} g^{vA}\label{exact} \eea and the following asymptotic Killing equations 
\bea
\label{approxLieuv}
       & \mathcal{L}_{\xi}g^{uv} = O\big(v^{-2}\big), \ 
        \label{approxLieuA}
        \mathcal{L}_{\xi}g^{uA} = O\big(v^{-3}\big), \ 
        \label{approxLievA}
        \mathcal{L}_{\xi}g^{vA} = O\big(v^{-3}\big), \nonumber \\
      &  \label{approxLiezz}
        \mathcal{L}_{\xi}g^{zz} = O\big(v^{-3}\big), \ 
        \label{approxLieww}
        \mathcal{L}_{\xi}g^{\zbr\zbr} = O\big(v^{-3}\big), \ 
        \label{approxLiezw}
        \mathcal{L}_{\xi}g^{z\zbr} = O\big(v^{-4}\big). \nonumber \\
\eea
After some calculation, this fixes the AKVs to be
\begin{subequations}\label{finalxi}
    \begin{align}
        \xi^u &= f+ \frac{1}{2}\, \alpha^A_3\, \partial_A f\, v^{-2} + \frac{1}{3}\, \alpha^A_4\, \partial_A f \, v^{-3}+O\big(v^{-4}\big) \\
        \xi^v &= -\frac{1}{2}\, D_A Y^A\, v+\big( T+\Delta_{\gamma} T\big)+O\big(v^{-1}\big) \\
        \xi^A &= Y^A-2\, \gamma^{AB}\, \partial_{B} f\, v^{-1}+O\big(v^{-2}\big)      
    \end{align}
\end{subequations}
with $f\equiv\xi_{0}^{u}=\psi(z,\zbr)\, u/2 + T(z,\zbr)$,  with $\psi(z,\zbr)= D_AY^A$ and $Y^A$ is a conformal Killing vector on the sphere with $Y^z\equiv Y^z(z), Y^{\zbr}\equiv Y^{\zbr}(\zbr)$. It should be immediately clear that $T$ captures supertranslations and the $Y^A$ are the (super-)rotations.  

The exact and asymptotic Killing conditions do not fix the sub-leading fall-offs in the AKVs uniquely in terms of supertranslations and super-rotations. In fact, while $\xi^u$ is completely fixed to all orders, we find three infinite towers of arbitrary functions of the angles $z$ and $\zbr$ in the coefficients of $\xi^A\equiv(\xi^z, \xi^{\zbr})$ and $\xi^v$. Remarkably, these are in one-to-one correspondence with the trivial integration ``constants" we encountered in the fall-offs. This makes it intuitively plausible that these trivial AKV coefficients have precisely enough freedom to set the corresponding trivial gauge parameters to zero, if we choose. By determining the action of the Killing vectors on the solution space \cite{Big}, one can explicitly show that this is indeed the case. 

Even though the calculations above are superficially different from those in Bondi gauge, the non-trivial AKVs that are left at the end contain supertranslations and super-rotations. This is a very strong suggestion that the asymptotic symmetries contain BMS. We can show this explicitly by computing the Barnich-Troessaert bracket of the AKVs, as defined in eqns (4.12) of \cite{Barnich}. We have checked that the Barnich-Troessaert bracket $[\xi_1,\xi_2]_M$ has leading behavior identical to that of \eqref{finalxi} with $\widehat{T}$ and $\widehat{Y}^A$ defined in (4.11) of \cite{Barnich} replacing $T$ and $Y^A$. It is also straightforward to check that it satisfies the exact Killing conditions \eqref{exact}. A more complete discussion of the asymptotic symmetry algebra, including the possible enhancement beyond BMS, will be presented elsewhere \cite{KP2}. 


Finally, we can also compute the covariant surface charges \cite{Iyer, Brandt}. This is usually omitted if one is not trying to compute the algebra of these charges (see eg. \cite{Barnich} as opposed to \cite{Barnich2}). But in our case, it is of some interest because computing them explicitly demonstrates that the trivial integration ``constants" (ie., all except the outliers mentioned in \cite{hardwire}) are indeed trivial gauge parameters -- they do not appear in the charges in Einstein gravity. We have checked this via direct calculation \cite{Big}.

\section{Asymptotic CPT Invariance}

We can repeat what we did for $\scrip$ to $\scrim$ by considering an expansion in $\frac{1}{u}$ with fall-off coefficients which are functions of $(v,z,\zbr)$, see footnote \cite{future-past}. In what follows, all corresponding functions (metric fall-off coefficients as well as AKV fall-off coefficients) in the future and past will be denoted by the same name but the past quantities will have a tilde (\ $\tilde{}$\ ). In order to define a well-defined scattering problem in flat space, Strominger \cite{Strominger1} suggested the matching between certain quantities in the future and past Bondi gauges. This ensures that the BMS frames in the past and future can be compared. 

A key observation for us, is that the quantities that are matched across $i^0$ by Strominger are the analogues of the integration ``constants" in our language. In particular they all arise as boundary values (at $u=-\infty$) of quantities that are $u$-integrals. In the Bondi case, these are the shear, mass aspect and angular momentum aspect at $\scrip_-$ (and $\scrim_+$). This suggests that a direct adaptation of this philosophy to our case can be accomplished if we find a systematic way to match  {\em all} our integration ``constants" (both trivial and physical) in the future and past. It turns out that a simple rationale for doing this exists, and that is to demand {\em asymptotic} CPT invariance for all solutions.  

(C)PT transformation in Minkowski space flips the signs of Cartesian coordinates and takes the form
\bea
u \rightarrow -v, \ v \rightarrow -u, \ z \rightarrow -\frac{1}{\zbr}, \  \zbr \rightarrow -\frac{1}{z}
\eea
in double null coordinates \eqref{Mink}. Inspired by this, we will demand asymptotic CPT invariance to mean order by order conditions relating the future and past fall-off coefficients of the gauge functions. We illustrate this using $\lambda_2$ by demanding (schematically):
\bea
\frac{\lambda_2(u \rightarrow -\infty, z, \zbr)}{(v\rightarrow \infty)^2} = \frac{\tilde\lambda_2(v \rightarrow \infty, -1/\zbr, -1/z)}{(u \rightarrow -\infty)^2}. \label{matching}
\eea
The solution to the $\lambda_2$ constraint -- which is second order, \eqref{lambda2eq} -- will be of the form
\bea
\lambda_2 =\lambda_2^0(z,\zbr) + u \ \lambda_2^1(z,\zbr)+\Lambda_2(u,z,\zbr) \label{newformp}
\eea
By demanding suitable fall-offs on shear/news or by renormalizing suitably (see the next section), we can ensure that $\Lambda$ vanishes at $u\rightarrow -\infty$. This implies via \eqref{matching} that we need matching conditions on the two integration ``constants" $\lambda_2^0$ and $\lambda_2^1$ of the form:
\bea
\lambda_2^0(z,\zbr)=\tilde \lambda_2^0\left(-\frac{1}{\zbr},-\frac{1}{z}\right), \ \lambda_2^1(z,\zbr)=-\tilde \lambda_2^1\left(-\frac{1}{\zbr},-\frac{1}{z}\right) \nonumber
\eea
In general, the matching condition is of the form
\bea
c(z,\zbr) = \pm \tilde c\left(-\frac{1}{\zbr}, -\frac{1}{z}\right) \label{anti}
\eea
Here $c$ stands for {\em any} of the integration ``constants" in the future direction and $\tilde c$ is its past counterpart. The sign is chosen to be positive if the sum of the powers of $u$ and $v$ that multiplies the ``constant" is even, and negative otherwise  -- except for a minor caveat in the case of $\alpha^A$. For $\alpha^A$ fall-off coefficients, there is an extra sign due to the fact that switching $(u,v) \leftrightarrow (-v,-u)$ requires an extra sign flip to bring our metric ansatz \eqref{doublenull} to its original form. In other words, $\lambda$ and $\Omega_{AB}$ transform in the trivial representation of the $\IZ_2$ associated to CPT, while $\alpha^A$ lives in the non-trivial representation. The AKV fall-off ``constants" can also be similarly matched by demanding that the matchings of the metric parameters above is respected by them \cite{Big}. The result is that we are left with a single copy of the BMS algebra, the diagonal BMS$^0$ of Strominger.

We have explicitly checked that the matching conditions are correct for the physical data defining Schwarzschild and Kerr in our gauge, by doing an asymptotic coordinate transformation from Bondi gauge to future and past SDN gauges. Let us also note that the explicit anti-podal inversion in the angular coordinates \eqref{anti} is invoked, because we are working with two copies of the {\em same} gauge in the future and past charts. The matching in the Bondi case is done between two manifestly {\em different} gauges, and the past angular coordinates are usually defined with respect to the South pole to (misleadingly) simplify notation. 

\section{Renormalization at $\scrip_-$ and $\scrim_+$}

As discussed, the constraints on the fall-off coefficients in $\lambda$ and $\alpha^{A}$ take the universal form 
\bea
\partial_u \beta(u,z,\zbr)& = B(u,z,\zbr), \\ 
{\rm or} \ \ \partial^{2}_u\ \mu(u,z,\zbr)& = M(u,z,\zbr), 
\label{schematic}
\eea 
i.e., they are derivative constraints on $u$. Generically they are first derivative constraints, these we denote collectively by $\beta$ above. The two exceptions are the second derivative conditions on $\lambda_2$ and $\alpha^A_4$, which are collectively denoted by $\mu$ in \eqref{schematic}. On top of these, there are also purely algebraic constraints that determine fall-off coefficients in $\Omega_{AB}$, but these are more familiar and therefore less interesting. In what follows we will discuss $\beta$, but very similar comments exist for $\mu$ as well, after two integrations \cite{Big}.

A natural prescription to solve the constraint is to write
\bea
\beta(u,z,\zbr)=b(z,\zbr)+\int_{-\infty}^u\  B(u',z,\zbr)\ du'.\label{lightray?}
\eea
where the lower end of the integral can be viewed as being at $\scrip_-$. In fact in the Bondi gauge, the mass aspect can be written as such an integral, and Christodoulou-Klainerman (CK) fall-offs \cite{CK1, CK2} on the news were invoked \cite{Strominger1} to argue that the integral converges at the lower end. Making a similar argument in the case of the angular momentum aspect requires more discussion \cite{Kapec, Strominger} because one finds an integral containing the shear itself (and not just the news) on the right hand side. 
In particular, it has been suggested that the matching conditions between the future and past may get anomalous corrections, see eg. footnote 29 of \cite{Strominger} and references therein.

The ubiquity of structures like \eqref{lightray?} in our set up suggests a natural paradigm for dealing with such questions, and allows the possibility of more general fall-offs than CK. The idea is to regulate spatial infinity, as is often done in discussions of the Ashtekar-Hansen gauge \cite{Asht}. This suggests in our case that we view these integrals as being cut off at a lower limit $u=-1/\epsilon$. We write 
\bea
\beta(u)&=&\beta^R+\int_{-1/\epsilon}^u\  B(u')\equiv\beta^R+\widehat{B}(u)-\widehat{B}(-1/\epsilon) \nonumber \\
&\equiv &\beta^B + \widehat{B} (u). \label{renorma}
\eea
where we have suppressed the angle dependence for brevity. This naturally suggests that the bare integration ``constant" $\beta^B$ should in fact be taken as cut-off dependent, and the matching should be done in terms of the renormalized (finite) data $\beta^R$. Equivalently, we can view the $\beta$ above as the analogue of the ``renormalized Lagrangian", and from the second line of \eqref{renorma} we see that when $u \rightarrow -\infty$ it has a (potential) divergence from $\widehat{B}(u)$, which cancels the (potential) divergence in $\beta^B$ \cite{renorm}. It is clear that $u$ (or perhaps $e^{-u}$) has features of an energy scale.


\section{Black Holes and Log Fall-Offs}

A feature of our fall-offs is that they do not involve logarithmic terms. They are smooth at infinity in terms of the {\em null} coordinates. Smoothness of null infinity is historically a complicated problem in general relativity, see eg. \cite{Kehr} for a recent discussion. More work is clearly needed to understand our result fully, here we will make one elementary comment about Schwarzschild black holes. 

The relation between the conventional (Eddington-Finkelstein) $U$ and $V$ coordinates and the Schwarzschild radial coordinate $r$ takes the well-known form (see eg., \cite{Schmidt}):
\bea
V-U=2r + 4 M \log (r/2M -1) \label{edd}
\eea
which at large-$V$ leads to
\bea
\frac{1}{r} = \frac{2}{V}+\frac{8M}{V^2}\, \ln V+\frac{2U}{V^2} + \cdots.\label{logf}
\eea 
This clearly contains log terms. One can write down Schwarzschild in these double null coordinates by thinking of $r$ as an implicit function \eqref{edd} of $U$ and $V$ and $t=\frac{V+U}{2}$. We will call it the Schmidt-Stewart coordinate system for Schwarzschild. It clearly satisfies our gauge conditions \eqref{dn} and \eqref{sdn} -- the difference is that the asymptotic expansion of the metric functions is no longer only in powers of $1/v$ but also contains (powers of) logs. 

One can understand the relationship between the two double null coordinates by doing an asymptotic coordinate transformation from Schwarzschild written in (future) Bondi gauge, to the SDN gauge. One can reach our SDN gauge by only invoking power law fall-offs in the coordinate transformation if one chooses. The integration ``constants" \cite{const} in our gauge are in one to one correspondence with the left-over freedom in the coordinate transformation after demanding \eqref{dn} and \eqref{sdn}. 

But one can also try asymptotic coordinate transformations that contain powers of logs. It turns out that the log terms also come with their own integration ``constants" and one can reach the Schmidt-Stewart coordinate system by choosing specific values for these integration ``constants". This may seem like an indication that the Schmidt-Stewart coordinates are a trivial gauge transformation (containing logs) away from our power law fall-offs. But there is a catch. The trouble is that in the Schmidt-Stewart coordinates one can compute the surface charges, and they turn out to be divergent if all the BMS AKVs are turned on \cite{Big, Schmidtlam}.  This suggests that Schmidt-Stewart coordinates lie in a different superselection sector, and also explains why we were able to retain familiar physics (like mass aspect) even though we only allowed power law fall-offs. 

The general message of these observations is that large gauge transformations (with divergent, finite or vanishing charges) may have interesting implications for discussions of analyticity at infinity.

\section{Outlook}
 
We have identified a new gauge for asymptotically flat spacetime that looks promising for discussions of holography. It differs from the extant gauges in that it can treat the future and past ($\scrip$ and $\scrim$) on an equal footing as would be natural if the $S$-matrix is the natural observable. Unlike conventional AdS/CFT \cite{Witten} there are {\em two} natural holographic directions and they are {\em null} instead of spacelike. We have identified a set of fall-offs and associated AKVs that can reproduce both the future and past BMS algebras. 
The diagonal BMS$^0$ symmetry emerges naturally upon demanding {\em asymptotic} CPT invariance. A natural notion of {\em renormalization} of asymptotic data emerges, and suggests the possibility of fall-offs more general than CK. Our fall-offs were smooth at infinity along a {\em null} coordinate, and yet we were able to retain familiar physics.  

Perhaps the most remarkable feature of the gauge is the presence of trivial gauge parameters and associated trivial diffeomorphisms. We feel that we have only scratched the surface of these trivial gauge parameters in this paper, and suspect that they have much more to say about the complete asymptotic symmetries of flat space quantum gravity. The known set of soft theorems suggests that the symmetries of flat space are much bigger than what is currently known, see \cite{StromingerW} for a very recent proposal. The fact that a handful of early integration ``constants" were {\em not} truly trivial, we find to be quite suggestive in this regard \cite{hyper}. Including them will clearly enhance the asymptotic symmetries by more arbitrary functions on the sphere, on top of supertranslations. We strongly suspect however that this is not the end of the story -- the whole tower of trivial gauge parameters may eventually play a role, in settings more general than classical Einstein gravity. We hope to come back to some of these questions in the near future. 

We conclude by including a (very incomplete) list of very recent works on celestial holography, as a point of entry into the subject \cite{refs}.

\section{Acknowledgments}

We thank Mrityunjay Nath for early collaboration and the organizers and participants (especially Alok Laddha and Amitabh Virmani) of ISM2021, Roorkee, for feedback on a talk presenting this material.



\begin{thebibliography}{99}
\bibitem{AshtHans} Although less numerous, there are also papers that aim for holography at spatial infinity (see eg., \cite{Marolf}), following Ashtekar and Hansen \cite{Asht}.
\bibitem{S} We are open to the possibility that a new approach to looking at (and perhaps going beyond) the S-matrix may be necessary. Some papers we find intriguing are \cite{S1, S2, S3}.
\bibitem{cod1} Codimension one surfaces are very natural for holographic purposes, both from black hole entropy expectations as well as well as considerations of diffeomorphism invariant observables, see \cite{ACD1} for some related discussions.
\bibitem{cod2}  In fact it is the naturalness of the ACD arguments in \cite{ACD1,ACD2} that made us suspect that a useful double null gauge for handling $\mathscr{I}^{+}$ and $\mathscr{I}^{-}$ may exist.
\bibitem{Weyl} When we demand Einstein equations together with matter dying down sufficiently fast at infinity, asymptotic Ricci flatness in some form is automatic. Then, fall-offs in Riemann translate to suitable fall-offs in Weyl.
\bibitem{lamD} Because algebraic constraints relate $\Omega_{AB}$  coefficients to those in $\lambda$ and $\alpha^A$, there is some freedom in what we call independent -- eg., there is an algebraic relation connecting $\mathcal{D}_{z\zbr}$ and $\lambda_2$. We will view the data appearing in $\lambda$ (and not in $\Omega_{z\zbr}$) as independent.
\bibitem{Subtlety} We will comment on the various caveats, as they arise. It has not escaped us that an infinite tower of arbitrary functions on the sphere is tantalizingly close to a huge enhancement of the BMS symmetry. But in Einstein gravity, we find that most of them have trivial asymptotic surface charges.  So we will be conservative in this letter.
\bibitem{RiemFlat} We will work with vacuum Einstein gravity in this letter and \cite{Big}, but we expect that closely related statements will hold even when there is matter.
\bibitem{Exceptions} There are two exceptions to this, namely $\lambda_2$ and $\alpha^A_4$, which have second order constraints. This is analogous to the fact that the mass and angular momentum aspect in Bondi gauge have differential constraints, even though all other data are purely algebraically determined.
\bibitem{newsdef} Note that we are working with a trivial conformal factor on the sphere in this letter.
\bibitem{shear} Supertranslations affect shear in the same way that they do in Bondi. The details of the action of AKVs on solutions can be found in \cite{Big}, but we will not need it here.
\bibitem{renorm} Recall that the bare Lagrangian contains finite parameters and needs counter-terms, while the renormalized Lagrangian contains cut-off dependent (divergent) parameters and leads to finite answers \cite{Ramond}.
\bibitem{const} Note that the ``constants" are true constants in Schwarzschild.
\bibitem{Schmidtlam} This can be seen quite simply. Along with the logs, Schmidt-Stewart coordinates also lead to a non-vanishing $\lambda_1$ term in the fall-offs. This immediately leads to divergences in the charges.
\bibitem{hyper} These ``hypertranslations" and ``hyperrotations" will be discussed elsewhere, and may be behind the sub-subleading soft graviton theorem discussed in the literature \cite{Cachazo}. See analogous discussions for supertranslations \cite{Prahar} and super-rotations \cite{Sabrina} which lead to the leading and subleading soft graviton theorems.
\end{thebibliography}
\end{document}